\documentstyle[12pt]{article}
\begin{document}
\begin{titlepage}
\begin{center}
{\Large\bf Contribution of intermediate stage gluons to $J/\Psi$
suppression in Lead-Lead collisions at 158AGeV}
\end{center}
\vspace{1cm}
\begin{center}
{\large  J\'an Ft\'a\v{c}nik${}^{a)}$, J\'an Pi\v{s}\'ut${}^{a)b)}$
 and Neva Pi\v{s}\'utov\'a${}^{a)}$}
\end{center}
\vspace{1cm}
\begin{center}
  {\it 
  a) Department of Physics,Comenius University,
  SK-84215 Bratislava,Slovakia \\
 b) Laboratoire de Physique Corpusculaire,Universit\'{e}
  Blaise Pascal,Clermont-Ferrand, F-63177 Aubi\`{e}re,Cedex, France }
\end{center}
\vspace{1cm}
\abstract
{We point out that dissociation of $J/\Psi $ by partons (mostly gluons)
present in the intermediate stage of heavy-ion collisions can explain 
  $J/\Psi $ suppression observed recently by the
NA-50 Collaboration at the CERN-SPS in Pb-Pb interactions. Suppression by
intermediate stage gluons represents an additional multiplicative factor
to that given by Gerschel- H\"{u}fner mechanism. The agreement with data
on $J/\Psi $ suppression both in light- ions induced nuclear collisions
and in Pb-Pb interactions requires that the life- time of intermediate
stage gluons increases with the nucleon numbers of colliding nuclei.

In our model the energy density of intermediate stage gluons in Pb-Pb
collisions approaches  for a short time the critical
density.} 

\vspace*{1cm}
{\bf Revised version, 15.Sept.1996}

{\bf Department of Physics,Comenius University,Bratislava}
\end{titlepage}
\section{Introduction}
\label{intro}
Suppression of $J/\Psi $ in heavy-ion collisions has been suggested as
a signature of Quark-Gluon Plasma by Matsui and Satz [1] a decade
ago. A sizeable suppression has been really observed by the NA-38
Collaboration [2] in collisions of Oxygen and Sulphur ions with heavy
nuclei.

As alternative explanations, suppression of $J/\Psi $
 by  thermalized hadron gas [3,4] or by nucleons present in nuclei [5]
have been suggested.New experimental data [6,7] on the A-dependence of
$J/\Psi$ production in hadron-nucleus collisions have shown that the
cross-section for $J/\Psi$ disintegration in a collision with a fast
nucleon is about 6mb. This has permitted to Gerschel and H\"{u}fner [5]
to describe satisfactorily the data on $J/\Psi$ suppression in hadron-
and light ion- collisions with heavy nuclei by disintegration of
$J/\Psi$ in collisions with nucleons.

The recent results of the NA-50 group show [8] that the $J/\Psi$
suppression in Pb-Pb collisions at laboratory energy of 158 GeV per
nucleon is by a factor of two stronger than that
given by the extrapolation of the
Gerschel-H\"{u}fner curve. Moreover, $J/\Psi $ suppression depends
rather strongly on centrality of Pb-Pb collision.

In this situation it is natural to search for a mechanism which would
be able to provide the additional $J/\Psi$ suppression in Pb-Pb without
increasing significantly the suppression in  collisions of lighter ions.

In early work [3,4] authors have studied $J/\Psi$ suppression by
a thermalized hadron gas. In view of recent studies by Kharzeev,Satz and
McLerran [9] it seems that cross-sections for the disintegration of
$J/\Psi$ by hadrons in a thermalized
hadron gas with temperatures of 150- 200MeV
 are rather low so one is
inclined to search the relevant mechanism in the intermediate
stage of heavy-ion collision. Out of possible candidates one can think
of at least two possibilities:

The former is the disintegration of $J/\Psi$ by fast
 secondary hadrons (mostly pions).
In a larger system, like Pb-Pb, some secondary
 pions have more time to complete their
formation [10] and are able to disintegrate $J/\Psi$. Fast pions
accompanying parent nucleons would effectively increase the cross-section
for inelastic nucleon-$J/\Psi$ collisions in the Gerschel-H\"{u}fner
picture.

The latter is the suppression of $J/\Psi$ by partons (mostly
gluons) produced in semi-hard nucleon-nucleon sub-collisions. As argued
in Ref.[11] and discussed below,
 the uncertainity principle leads to a relative suppression
of soft interactions in nucleon-nucleon sub-collisions in  interactions
of heavy ions at high energies.
 Secondary partons produced in one sub-collision 
interact with $J/\Psi$ produced in another sub-collision and this leads
to additional $J/\Psi$ suppression.

The purpose of the present note is to study in some detail the latter
mechanism, to estimate the corresponding $J/\Psi$ suppression and to
find conditions under which this additional suppression gives
a factor of about two in Pb-Pb collisions, and 
factor close to one in interactions of lighter ions.

Suppression of $J/\Psi $ by gluons has already been discussed by
Wong [12] and by Xu et al. [13]. In contradistinction to Wong [12]
we emphasize here that the time during which intermediate stage gluons
exist prior to their hadronization increases with increasing gluon
density and as a consequence of that also with increasing nucleon
numbers of colliding nuclei. According to the view we are advocating
here, this effect is responsible for the anomalous $J/\Psi $
suppression in Pb-Pb collisions as observed by NA-50 Collaboration [8].

Xu et al.[13] consider an equilibrating parton plasma which, starting
from the original inequilibrium approaches thermal equilibrium. Such
a process extends over the time of several fm/c and might correspond
to what will happen in the RHIC and LHC energy range. In our model
we assume that intermediate stage gluons are present in the system only
during 0,5 - 1.5 fm/c and after that they hadronize without approaching
an equilibrium in the partonic stage.

We also put emphasis on the relationship between this picture of 
$J/\Psi $ suppression and the enhanced production of strangeness in
heavy- ion collisions.

The paper is organized as follows: In the next Section we shall briefly
describe the picture of the space-time evolution of heavy- ion collision
we are using [11]. In Sect.3 we study the time-evolution of density
of semihard gluons. Sect.4 contains estimates of $J/\Psi$ suppression by
this mechanism. Comments and conclusions are presented in Sect.5.
\section{Parton model of heavy-ion collisions in the 200AGeV energy
 range}
\label{model}
According to models [14] based on Perturbative Quantum Chromodynamics
(PQCD) of proton-proton interactions at high energies
 the collision consists of two
parts: a semi-hard parton - parton interaction (mostly gluon - gluon; in
what follows we shall refer explicitly to gluons) populating the central
rapidity
region and fragmentation of "wounded" nucleons [15] (less
 understood, because
not described by PQCD). Models of semi-hard collisions introduce a cut-off
on momenta of participating gluons $p_T > p_T^0 $ in order to have finite
cross- sections and not to enter the non- perturbative region. The value
of $p_T^0 $ depends somewhat on the energy of $pp $ collision and is in
the region $0.5 GeV < p_T^0 < 1 GeV $ for
 lab. energies of a few hundred GeV.
 The existence of the cut-off may be connected with the screening
of individual partons by other partons in the system.

The presence of two mechanisms of hadron production in $pp$ collisions is
indicated by data on strange baryon production [16] where $\bar\Lambda $
seems to be produced  by the harder mechanism populating
the region $ -0.3 < x_{cm} < 0.3 $, whereas $\Lambda $ is produced by both
mechanisms, the former being the semi-hard and the latter due to proton
fragmentation. These two mechanisms are visible also from
data on transverse momentum spectra of $\Lambda $ and $\bar\Lambda $. The
harder mechanism is connected with larger transverse momenta and has
a smaller $p_T$ -slope. The slope of $\Lambda $ production has a break in
the slope parameter, corresponding to the transition from the softer to
the harder dynamics. This break of the $p_T$ -slope of $\Lambda $-production
has been observed also at lower energies of 12,4 GeV/c and 200 GeV/c [17]
what indicates the presence of both mechanisms also at lower energies. The
slope of $\bar\Lambda $ is smaller and about the same as the slope of
$\Lambda $ at higher $p_T $.
Similar findings on the spectra of $\Lambda $ and $\bar\Lambda $ has been
reported also in the study of strange particle production at 360 GeV/c [18].
For a detailed review of data and references on strange particle production
see Kachelhoffer and Geist [19].

These features of data lead us to conjecture that in
nucleon-nucleon collisions at lab.energy of about 200 GeV, there are two
mechanisms present:

(i) a harder one, probably due to semi-hard gluon- gluon interaction,
populating central $x_{cm}$ region,

(ii) a softer one, presumably associated with nucleon fragmentation,
populating the beam and target proton fragmentation regions and
extending partly to the central region. Near $x_{cm} \approx 0 $,
the contribution from (i) is presumably dominating as indicated by
roughly equal magnitudes of cross-sections for $\Lambda $ and
$\bar\Lambda $ at $x_{cm} \approx 0 $ [14].

 Data on $\Lambda $ and $\bar\Lambda $ production thus
give evidence on the presence of the two mechanisms at $E_{lab} \approx
200 GeV$ and indicate that semi-hard PQCD gluon-gluon interactions
combined with fragmentation of wounded nucleons might provide a useful
picture of what happens in first stages of both
  pp and nucleus- nucleus interactions in this energy range. Note that
this picture does not contradict the work  [20] on nuclear stopping
power at $E_{lab} \approx 100 GeV$ in the interpretation of
Dat\'{e} et al.[20].

In ion- ion interactions  soft exchanges
 in nucleon- nucleon sub- collisions
are suppressed by the uncertainity principle, the argument being similar
to that of the Landau- Pomeranchuk mechanism, for a review see Ref.[21].
In  heavy- ion collision each nucleon collides  with a few
nucleons from another ion. If time and longitudinal distance
 between two subsequent collisions are
 $\Delta t$ and $\Delta z$ the uncertainity principle
 mechanism suppresses processes with
 energy and longitudinal momentum transfer 
\begin{equation}
\Delta p_z < {\frac {\hbar }{\Delta z}},\qquad \Delta E < {\frac {\hbar}
{\Delta t}}
\label{eq1}
\end{equation}
Mean free path for a nucleon passing through a nucleus at rest is about
3fm. When considering a collision of two ions at $E_{lab} \approx $
 200 GeV,
in the c.m.s. of nucleon -nucleon collision both nuclei are
contracted  by the Lorentz factor $\gamma \approx 10$, the mean free path
becomes shorter by the factor $\gamma$ and Eq.(1) leads to a suppression
of energy and momentum transfer with
\begin{equation}
\Delta p_z < 600MeV/c ,\qquad \Delta E < 600MeV
\label{eq2}
\end{equation}
Just above these limits one is at the border of applicability of
the notion of semi-hard collision between partons and accurate calculations
are not very reliable.
 In spite of that, due to hints provided by data on
$\Lambda $ and $\bar\Lambda $ production discussed above, we suppose that
the picture based on semihard production of gluons (which hadronize later
on) and fragmentation of nucleon remnants may provide a useful insight
into the first stages of ion-ion collisions even in the energy range of
$E_{lab} \approx $ 200 AGeV . This picture is also  relevant for the
question of $J/\Psi$ suppresion. Gluons produced in nucleon- nucleon
sub- collisions can collide
 with $J/\Psi $ and disintegrate it. According to
Ref.[9] the inelastic $g + J/\Psi$ cross-section reaches its maximal
value around 3mb for gluons with momentum of about 1GeV/c and this is
 what one expects for gluons originated in semi-hard interactions and
$J/\Psi$ at rest in the c.m.s. of nucleon- nucleon system.

In the next Sect. we shall discuss the
expected evolution of intermediate stage gluon densities
 in heavy-ion collisions.
\section{Evolution of gluon densities in heavy-ion collisions in the
 200 AGeV energy range}
\label{evolution}
For the sake of simplicity we shall picture the colliding nuclei A and B
 as cyllinders
of lengths $2L_A$ and $2L_B$ and radii $r_A$ and $r_B$. These
 parameters are
fixed for each nucleus by requiring that the volume of the cyllinder and
the volume of the sphere are the same and that the value of $< z^2>$ where
$z$ is the distance from the centre of the sphere (from the centre of the
cyllinder) along the z-axis, identical with the axis of rotational
symmetry for the cyllinder. In this way we find
\begin{equation}
L_A = {\sqrt {3\over 5}}R_A ,\qquad
 r_A^2= {2\over 3}{\sqrt {5\over 3}} R_A^2
\label{eq3}
\end{equation}
Space-time evolution of two colliding rows of nucleons, each row consisting
of a tube along the z-axis, can be visualized simply as
 shown in Fig.1, where
one can see the position of both rows at any value of time. Taking a
particular value of z and making in that point straight line parallel to
the t-axis we find the time interval during which  nucleons of both
rows collide at this value of z. The time when nucleons start to collide
in point z is denoted as $t_1(z)$ and the time when collisions cease as
$t_2(z)$.The rate ${(dn/dt)}_0$ at which gluons are produced in the point
z vanishes for $t\le t_1(z)$ as well as for $t\ge t_2(z)$. Within the
time interval $t_1(z) < t < t_2(z)$ the rate is given as
\begin{equation}
\alpha \equiv {({dn\over dt})_0}=v_{rel}\gamma {\rho}_0 \gamma {\rho }_0
\sigma 2C
\label{eq4}
\end{equation}
where $v_{rel} \approx 2c$ is the relative velocity of colliding nucleons,
$\sigma $ is the cross-section for semi-hard nucleon-nucleon collision
leading to production of two gluons with momenta larger than about 0.5GeV,
${\rho }_0$ is the nucleon density in a nucleus at rest and
factor $2$ stands for the production of two
 gluons in one sub- collision. The factor $C$ takes into
account that  partons which are responsible
 for most of semi-hard interactions
are not Lorentz contracted. Positions of collisions within the nucleus
A will not be distributed within the distance $2L_A/\gamma $ but over a
larger distance $(2L_A/\gamma + \Delta )$ where $\Delta $ is about 1fm. The
correction factor C thus becomes
\begin{equation}
C={\frac {R_A}{R_A +\delta}}.{\frac {R_B}{R_B + \delta}}
\label{eq5}
\end{equation}
where
\begin{equation}
\delta ={1\over 2} \sqrt {5\over 3} \gamma \Delta \approx 6,5 \Delta
\label{eq6}
\end{equation}
  Note that this correction leads to rather lower densities for collisions
of lighter ions. To get a feeling for the numbers, consider Pb-Pb
collision with the following parameters: $v_{rel} \approx 2c$, $\gamma
\approx 10$ (this corresponds to $E_{lab} \approx $200 AGeV), $\sigma
\approx 20mb = 2fm^2$, ${\rho }_0 \approx 0,15 fm^{-3}$, $\Delta \approx
1fm$, $R_{Pb} \approx 7fm$. The resulting ${(dn/dt)}_0$ becomes about
$(4,8 gluons)fm^{-3}(fm/c)^{-1}$. For an S-S collision due
to the C-factor the corresponding $(dn/dt)_0$ is only about a half
of the value for Pb-Pb.

By estimating $\sigma \approx 20mb$ we make the crucial assumption
that a large
part of nucleon- nucleon interaction in ion- ion collision is due to
semi- hard interactions even at $E_{lab} \approx $200 AGeV. For this
assumption the role of principle of uncertainity, see
 Eqs.(1) and (2), is essential.

The second decisive parameter is the time $\tau $ which a gluon spends in
the system as a gluon. This parameter can be estimated as follows.
In  proton-proton collision one can imagine that the two gluons
separate, extend a colour string between themselves, the tension of
the string being  1GeV/fm, and after a time of about 0.5 fm/c, when the
energy of gluons is converted to that of the string tension,
 the hadronization starts. In a heavy-ion collision the situation
might be rather different. The region in which nucleons collide contains
coloured partons what screens  interactions between gluons.
Gluons produced in one of semi-hard interactions may interact with
softer  gluons of
incoming nucleons what would increase their number. They can also interact
with gluons originated by other semi-hard collisions what makes their
lifetime longer. We shall lump all these effects into a single
parameter $\tau $ and consider its value as a free parameter with a value
of about 0.5-1.5 fm/c, depending on the gluon density and on the size of
the intermediate gluonic system.

We assume that the density of gluons increases due to
semi- hard collisions in the region of
nucleon- nucleon collisions shown in Fig.1 and that this density is
decreasing due to the hadronization of gluons and their escape from this
region. Characterizing the loss of gluons by a characteristic time
$\tau $ we can write the equation for the time evolution of gluon
density in point z as:
\begin{equation}
\frac {dn}{dt} ={({\frac {dn}{dt}})_0} - {\lambda }n,\qquad \lambda =
{1\over \tau }
\label{eq7}
\end{equation}
with $\alpha \equiv (dn/dt)_0$ given by Eq.(4). The solution of Eq.(7) is
\begin{equation}
n(t)=\tau \alpha [1-e^{-\lambda (t-t_1)}]
\label{eq8}
\end{equation}
for
$$t_1\equiv t_1(z) \le t \le t_2 \equiv t_2(z) $$
and
\begin{equation}
n(t)=\tau \alpha [1-e^{-\lambda (t_2 - t_1)}]e^{-\bar\lambda (t-t_2)},
\qquad t\ge t_2
\label{eq9}
\end{equation}
where $\bar\lambda = 1/\bar\tau $ describes the rate at which gluons
disappear after nucleons ceased to interact in the point z.
 In making the
estimates we shall first assume that $\bar\tau = 0$, what corresponds to
$n(t)=0$ for $t>t_2$. Note,however,that $\bar\tau $ may become large when
the density of gluons approaches or excedes the critical density
corresponding to the phase transition to QGP. We shall return to this
point below.

The time dependence of gluon density as given by Eq.(8) has two simple
limiting cases:

(i) The time $\tau $ is small with respect to $t_2-t_1$. In that situation
the expression $\lambda (t-t_2)$ is large for most of time t within
$(t_1,t_2)$ and we have a constant gluon density
\begin{equation}
n(t) \approx \tau \alpha,\qquad t_1\le t \le t_2
\label{eq10}
\end{equation}
and
$$ n(t)=0 \qquad for \qquad t\le t_1, t \ge t_2 $$
This approximation overestimates the gluon density
for values of $z$ where $t_2(z)-t_1(z)$
is smaller or roughly equal to $\tau $ and underestimates the density
for situations when $t_2(z)-t_1(z)$ is larger than $\tau $ or when
$t>t_2(z)$. 

(ii) The time $\tau $ is large with respect to $t_2 -t_1$. In this case
gluon density is increasing linearly with time
\begin{equation}
n(t) \approx \alpha (t-t_1),\qquad t_1 \le t \le t_2
\label{eq11}
\end{equation}
\section{Suppression by intermediate stage gluons}
\label{suppression}
When the time dependence of gluon density is known we can calculate
the survival probability of $J/\Psi$ by standard procedures, see e.g.
Ref.[4]. For $J/\Psi $ at rest in the c.m.s. of nucleon- nucleon collision
the survival probability S due to interactions with gluons is given as
\begin{equation}
 S=\left\langle exp(-\int <{\sigma }_d> n(t') v_{rel} dt' \right\rangle
\label{eq12}
\end{equation}
where $<{\sigma }_d>$ is the mean value of the cross-section for $J/\Psi$
disintegration in $g+J/\Psi$ collisions obtained by averaging over momenta
of gluons. Most of gluons produced in semi-hard nucleon- nucleon collisions
 have momenta within the range $0.5GeV/c \le p \le 1.5GeV/c$. According to
estimates of the disintegration cross- section for $g+J/\Psi $ collisions
given by Kharzeev and Satz [9] we shall take
 $<{\sigma }_d > =2mb=0,2fm^2$. For
massles gluons $v_{rel}=c$.

We shall now calculate the $J/\Psi $ survival probability, first by using
the approximation in Eq.(10) and then by using the numerical 
solution of Eqs.(8,9).
 In the limiting case of Eq.(10) with constant
gluon density, Eq.(12) simplifies to
\begin{equation}
S=\left\langle exp(-\int a.dt)\right\rangle,
 \qquad a=<{\sigma }_d> \tau {(dn/dt)}_0 c
=const.
\label{eq13}
\end{equation}
 Averaging in Eq.(13) goes over all positions (z,t)
 in which $J/\Psi $ can be created and the integral in Eq.(13) has as the
lower limit the time $t_1(z)$ and
 as the upper limit $t_2(z)$ (when one assumes, as
we now do, that the density of gluons vanishes
 for $t>t_2(z)$). The Eq.(13)
can be then rewritten as
\begin{equation}
S={1\over P}\int dz \int dt e^{-a(t_2(z)-t_1(z))}
\label{eq14}
\end{equation}
where one integrates over the region of nucleon- nucleon collisions
shown in Fig.1. The "surface" of this region is denoted as $P$. The
integral in Eq.(14) can be explicitly performed and we find
for Pb-Pb collision
\begin{equation}
S={\frac {2}{(aL)^2}}[aL - 1 + e^{-aL}]
\label{eq15}
\end{equation}
where $L=2L_{Pb}/\gamma +\Delta $ as discussed between Eqs.(4) and (5).
Using Eq.(3) to calculate $L_{Pb}$ with $R_{Pb} \approx 7fm $ and taking
$\Delta =1fm$ we find $L \approx 2.2fm$.
 Value of $a$ is determined by using
Eq.(13) with the following parameters: ${<\sigma >}_d=2fm^2$, $\tau =0.5
fm/c$, $(dn/dt)_0 =4.8 fm^{-4}c$ as estimated below Eq.(6). In this way we
get $a=0.48fm^{-1}$. For the product $aL$ this leads to $aL=1,06$.
Inserting that into Eq.(15) we obtain
\begin{equation}
S_{Pb-Pb} (gluons) = 0.71
\label{eq16}
\end{equation}
Proceeding in the same way in the case of S-S collisions we get $L=1.6fm$,
$a=0.24fm^{-1}$ and $aL=0.38$. Inserting that into Eq.(15) leads to
\begin{equation}
S_{S-S} (gluons) = 0.88
\label{eq17}
\end{equation}
The results as given by Eqs(16) and (17) are not satisfactory, since
the additional suppression given by intermediate stage gluons is only
0.71 for Pb-Pb interactions but it is as high as 0.88 for S-S
interactions. The data would rather require about 0.5 for Pb-Pb
and something closer to one for S-S collisions. Since some errors
may be due to the approximation itself, we shall in what follows, use
numerical results based on solutions given in Eqs.(8) and (9). Survival
probability is calculated by Eq.(12). Some results are summarized
in Table 1. As can be seen from the Table there is no combination
of relaxation times $\tau $ and $\bar \tau $ which would give the
required patterns of $J/\Psi $ suppression for both S-S and Pb-Pb
interactions.

The required patterns can be obtained provided that we assume that for
a  system like Pb- Pb where
the dimensions and the gluon density are larger the relaxation time is
also larger. This is certainly not very surprising since we expect that
the relaxation of intermediate stage gluons has
 some  similarity with  a  diffusion process and the coefficient of
diffusion is proportional to the mean free path which decreases with
decreasing density of particles. 

As seen in Table 1 combinations with $\tau \approx 0.5 $ fm/c  and
$\bar \tau \approx $ 1.5 fm/c and $\tau \approx $ 1 fm/c and
$\bar \tau \approx $ 1 fm/c
 give additional suppression of $J/\Psi $
in Pb- Pb collisions of about 0.56.
 Combinations with $\tau \approx $
0.5 fm/c and $\bar \tau \approx $ 0.5 fm/c keep $J/\Psi $ suppression
at acceptable levels of S(S-S)$\approx $ 0.89 and S(S-Pb)$\approx $
0.83 in S-S and S-Pb collisions.
\section{Comments and conclusions}
\label{conclusions}
The approach we have used in estimating the evolution of gluons
produced in semi- hard nucleon- nucleon interactions is admittedly
oversimplified, but it permits rough estimates of $J/\Psi$ suppression
by these gluons. The process itself 
is rather short and is followed by hadronization of gluons present
in the intermediate stage. To get
a better understanding of the process one should study the behaviour
of the intermediate systems of gluons and find the dependence of
basic parameters,in particular of $\tau $ on the size of the system.
Such a dependence is able to explain the abruptness of the decrease of
survival probability of $J/\Psi $ as observed by the NA-50
Collaboration. The model we have used above explains the abrupt increase
of $J/\Psi $ suppression by an increase of the relaxation time of
intermediate stage gluons in Pb-Pb interactions. Gluons in the
intermediate stage are not supposed to approach the equilibrium
in the QGP stage and hadronize before such a state can be reached.
The system of hadrons may thermalize depending on the density and
size of the hadronic system.  

In spite of the crudeness of approximations made above we shall now
discuss what might be the relation of the system of gluons produced
in semi- hard nucleon- nucleon collisions to QGP.
The energy density
 ${\varepsilon }_g $, of these intermediate stage gluons
 can be simply estimated as
$$ {\varepsilon }_g \approx <\varepsilon > \tau (dn/dt)_0 \approx
<\varepsilon > 2.4{fm}^{-3} $$
where $<\varepsilon > $ is the average energy of a semi-hard gluon.

Taking $0.5GeV \le <\varepsilon > \le 1GeV $ we obtain
$$ 1.2 GeVfm^{-3} \le {\varepsilon }_g \le 2.4 GeVfm^{-3} $$
According to lattice calculations [20] the critical temperature for
the phase transition to QGP is between 150 MeV and 200MeV,the lower
values being preferred.The corresponding energy densities are
$$0.8 GeVfm^{-3} \le {\varepsilon}_{crit} \le 2.5 GeVfm^{-3}$$
 The comparison of
these two regions of energy density  indicates
that the effect
observed by the NA-50 Collaboration [8]
might be connected with the closeness
of the energy density of semi- hard gluons to the critical one.
In our model it is natural to 
 assume that the relaxation time of the gluonic system
increases  substantially when the density  of gluons approaches the
critical one. The abrupt increase of $J/\Psi $ suppression could mean
that the NA-50 Collaboration have "touched"
the plasma and the experiments at RHIC and LHC will study the plasma
in detail.

In analyses [3,4] of the early NA-38 data [2] one was looking for a "kink"
in the dependence of $J/\Psi $ survival on the energy density as estimated
by the Bjorken formula (for a discussion along these lines see e.g. the
second paper in Ref.[4]). Even without a deeper theoretical understanding
a possible "kink" has been considered to be a sign of a
"threshold" signalling a transition to a different
dynamical regime of heavy-ion collisions. Increased accuracy of data has
shown [5] that up to S-U collisions there were no kinks,
also no thresholds and presumably also no transition.
 New phenomenon observed by the
NA-50 Collaboration has all the features one can expect from a kink
signalling a transition to a new dynamical regime.

The interpretation of results of NA-50 Collaboration as an indication
of a transition to another dynamical regime are further corroborated by the
recent data of NA-44 Collaboration at CERN-SPS [23] which give for the
longitudinal radius $R_L$ as obtained from the study of Bose-Einstein
correlations the value of about 6fm. This is much
 larger than  the value
found for S-Pb collisions [23,24,25]. 
Both phenomena deserve a careful and systematic
analysis, including a study of
 possible alternative explanations.
Only after having
excluded such alternative possibilities one could draw stronger conclusions
on the nature of the transition to a different dynamical regime in central
Pb-Pb collisions. If
 the interpretation presented above has some contact with
reality one can expect strong "threshold effects"
connected with some form of a kink also in the behaviour of
other signatures and perhaps also in shifts of masses of some resonances
[22,26].

The explanation of increased $J/\Psi $ suppression by increased life- time
of the intermediate gluonic system is quite natural and in a sense trivial.
What is non- trivial is the fact that the additional suppression can be 
obtained with expected values of cross- sections for both semi- hard
gluon- gluon collisions and disintegration of $J/\Psi $ in a collision with
a gluon. Non- trivial are also consequences
of this picture for other signatures of QGP.

The understanding of what is going on in heavy- ion collisions will most
likely come not from a single piece of data but rather via  consistency
of one picture with numerous types of experimental information. The picture
presented above leads to the following qualitative consequences. In
contradistinction to QGP followed by the mixed phase,
 the intermediate gluonic stage produces practically no
photons and no dileptons, its contributions to the life- time and
the size of longitudinal expansion are also smaller than those of
equilibrated QGP. For photon and dilepton production and for the
longitudinal radius $R_L$ one expect in this picture essentially only
the contribution from the thermalized hadron gas stage.

Strangeness is expected to be enhanced first by semi- hard gluon- gluon
collisions which are flavour blind and have probably enough energy to
produce more of heavier  strange particles than softer fragmentation.

The initial conditions for the hydrodynamical evolution
in this picture correspond more
to the Landau scenario than to the Bjorken one.

The intermediate parton stage is present in numerous models of $e^+e^-$
and hadronic collisions [27,28]. Our picture of heavy- ion collisions 
goes in a similar way to systems with higher density of gluons.

 The NA-50 data can be also described as $J/\Psi $
disintegration by comoving hadrons [29]. The two picture might differ
significantly in the question of strangeness enhancement, provided that
the expansion of the hadronic system is not long enough to establish the
chemical equilibrium by collisions of secondary hadrons.

Blaizot and Ollitrault [30], see also Kharzeev [31],
 have explored a scenario in which all 
$J/\Psi $  are totally suppressed in regions where the energy density
exceeds a critical value slightly greater than
that attained in S+U collisions.
Such a scenario indicates the formation of QGP and differs from the
scenario with shorter stage of intermediate gluons by the expansion
patterns.

In conclusion, we have suggested a picture of heavy- ion collisions
in which secondary particles appear either from fragmentation of 
nucleon remnants or from the evolution of intermediate system of gluons
formed by semi- hard gluon- gluon collisions. In this picture $J/\Psi $
is suppressed both by the Gerschel- H\"{u}fner mechanism and by the
intermediate system of gluons. The picture is able to describe the NA-50
data on Pb+Pb collisions provided that the life- time of the intermediate
gluon stage is increasing with  increasing nucleon numbers
of colliding ions and with increasing centrality of collision. For
central Pb+Pb
collisions the required life- time of the intermediate gluonic system
is about 1.5 fm/c, and the corresponding
  density of gluons in these collisions is close to the critical one.

{\bf Acknowledgements} We are indebted to F.Antinori, K.Geiger, W.Geist,
 R.Lietava, E.Quercigh, K.\v{S}afa\v{r}ik, L.\v{S}\'andor and P.Z\'avada
for numerous useful discussions on the problem of strangeness enhancement
in ion-ion collisions and to J.Castor, F.Devaux,
 J.-J.Dugne, P.Force, S.Louise,
J.-F.Mathiot, V.Morenas, G.Roche, and P.Saturnini
 for discussions on
 $J/\Psi$ suppression. We have particularly appreciated G.Roche's
insistence on the necessity to understand the rapid decrease of $J/\Psi$
survival probability.We are grateful to H.Meunier for valuable advices.
 One of the authors (J.P.) is indebted to G.Veneziano
for a hospitality at CERN-TH when a part of the work on strangeness
enhancement has been done and to B.Michel and G.Roche for the hospitality
at the Laboratoire de Physique Corpusculaire
 of the Blaise Pascal University,
 Clermont-Ferrand.
\vfill\eject
\begin{table}
\begin{center}
\caption {Table 1. The dependence of $J/\Psi$ survival probabilities on
      relaxation times $\tau$ and $\bar{\tau}$}
\label{blb}
\vskip 3mm
\begin{tabular}{|c|c|c|c|c|}
\hline
$\tau[fm/c]$ & $\bar{\tau}[fm/c]$ & $S(S+S)$ & $S(S+Pb)$ & $S(Pb+Pb)$ \\
\hline
0.3 & 0.3 & 0.93 & 0.89 & 0.82 \\
0.5 & 0.5 & 0.89 & 0.83 & 0.73 \\
0.5 & 1.0 & 0.83 & 0.76 & 0.64 \\
1.0 & 0.5 & 0.86 & 0.78 & 0.67 \\
1.0 & 1.0 & 0.79 & 0.70 & 0.56 \\
0.5 & 1.5 & 0.79 & 0.69 & 0.56 \\
1.0 & 1.5 & 0.74 & 0.62 & 0.48 \\
0.5 & 2.0 & 0.74 & 0.63 & 0.50 \\
1.0 & 2.0 & 0.68 & 0.56 & 0.42 \\
2.0 & 2.0 & 0.65 & 0.52 & 0.37 \\
\hline
\end{tabular}
\end{center}
\end{table}

Additional survival provability due to intermediate stage gluons is
calculated by using: Eqs.(8),(9) and (12) with $\sigma_g=2 fm^2$
(gluon pair production in a nucleon - nucleon collision) and
$\sigma_d=0.2 fm^2$ (disintegration of $J/\Psi$ in collision with a gluon).
\hfill
\vspace{2cm}

{\bf Figure Caption}

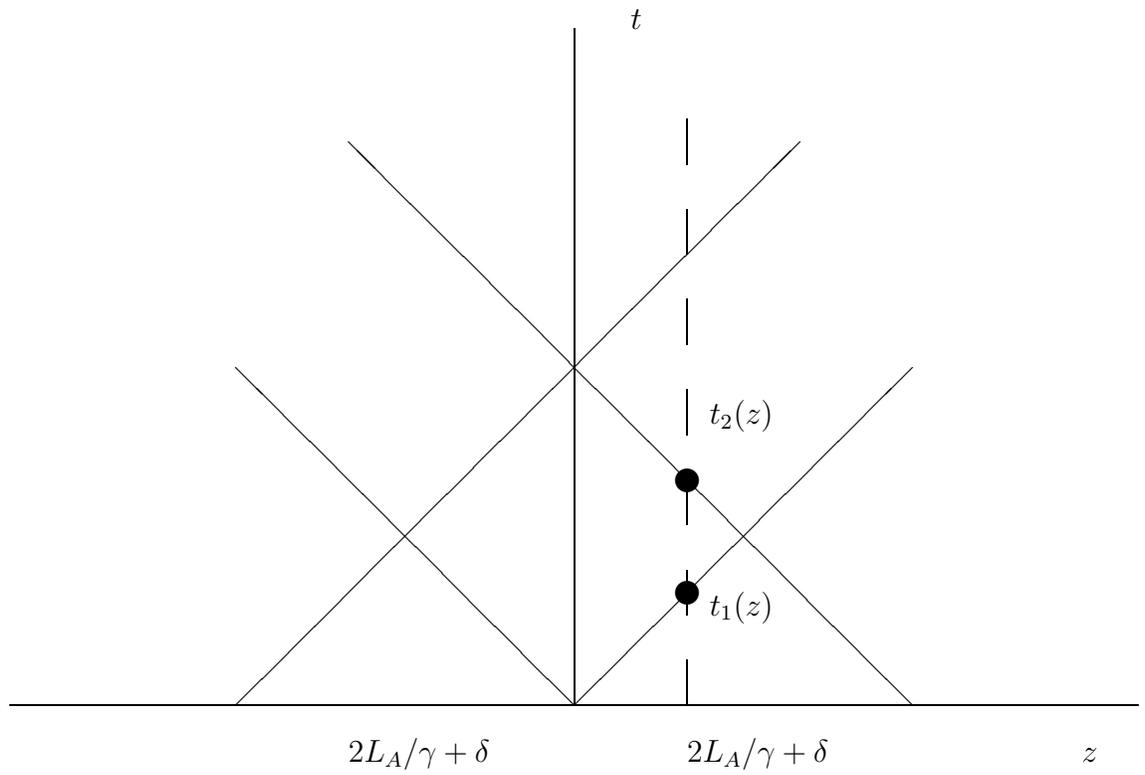
{\bf Fig.1} The z-t diagram of the space-time evolution of nucleon-
nucleon collision.Nucleons of two equal mass nuclei,modelled as cylinders,
collide within the square indicated.At a particular value of z the
nucleon- nucleon collisions start at $t=t_1(z)$ and end at
 $t=t_2(z)$. Length of each cyllinder is $2L_A/{\gamma} +\delta $ and
the symbols used are explained in the text.
\hfill
\vfill\eject

\vfill \eject
\setlength{\unitlength}{1.5cm}
\begin{picture}(10,7)
\put(0,1){\line(1,0){10}}
\put(5,1){\line(0,1){6}}
\put(2,1){\line(1,1){5}}
\put(5,1){\line(1,1){3}}
\put(5,1){\line(-1,1){3}}
\put(8,1){\line(-1,1){5}}
\multiput(6,1)(0,.8){7}{\line(0,1){.4}}
\put(6,2){\circle*{.2}}
\put(6,3){\circle*{.2}}
\put(6.2,1.8){$t_1(z)$}
\put(6.2,3.5){$t_2(z)$}
\put(3.0,.5){$2L_A/\gamma +\delta$}
\put(6.0,.5){$2L_A/\gamma +\delta$}
\put(9.5,.5){$z$}
\put(5.5,7){$t$}
\put(5,0){Fig.1}
\end{picture}


\begin{thebibliography}{23}
\bibitem{r1} T.Matsui and H.Satz, Phys.Lett. {\bf B178} (1986) 416
\bibitem{r2} C.Baglin et al.,Phys.Lett. {\bf B220} (1989) 471;
   C.Baglin et al., Phys.Lett. {\bf B255} (1991) 459; C.Baglin et al.,
   Phys.Lett. {\bf B270} (1991) 105; C.Baglin et al.,Phys. Lett. {\bf 345}
   (1995) 617
\bibitem{r3} A.Capella et al., Phys.Lett. {\bf B206} (1988) 354;J.-P.
 Blaizot and J.-Y. Ollitrault, Phys.Lett. {\bf B217} (1989) 386;
 S.Gavin, M.Gyulassy and A.Jackson, Phys.Lett. {\bf B207} (1988) 257;
 R.Vogt,M.Prakash,P.Koch and T.H.Hansson, Phys.Lett. {\bf B207} (1988) 263;
 R.Vogt, S.J. Brodsky and P.Hoyer, Nucl.Phys. {\bf B360} (1991) 67
\bibitem{r4} J.Ft\'a\v{c}nik,P.Lichard and J.Pi\v{s}\'{u}t, Phys.Lett.
 {\bf B207} (1988) 194; J.Ft\'a\v{c}nik et al., Zeit. f. Phys. {\bf C42}
 (1989) 139
\bibitem{r5} C.Gerschel and J.H\"{u}fner,Phys.Lett. {\bf B207} (1988) 253;
 C.Gerschel and J.H\"{u}fner, Zeit. f. Phys. {\bf C56} (1992) 171;
 C.Gerschel, Nucl.Phys. {\bf A583} (1995) 643
\bibitem{r6} S.Katsanevas et al.,Phys.Rev. {\bf 60} (1988) 2121
\bibitem{r7} D.M.Alde et al., Phys.Rev.Lett. {\bf 66} (1991) 2285
\bibitem{r8} P.Bordalo for NA-50 Collaboration, $J/\Psi$ suppression
 in Pb-Pb interactions at 158 GeV/nucleon, Talk at Rencontres de Moriond;
  M.Gonin, Talk at Quark Matter '96 Conference, May 1996, Heidelberg
\bibitem{r9} D.Kharzeev and H.Satz, Colour deconfinement and Quarkonium
 dissociation,CERN-TH/95-117,May 1995,To appear in Quark-Gluon Plasma II,
 R.C.Hwa (Ed.),World Scientific,Singapore;
 March,1996; D.Kharzeev and H. Satz, Phys.Lett. {\bf B334} (1994) 155,
 D.Kharzeev, L. McLerran and H. Satz, Phys.Lett. {\bf B356} (1995) 349,
 D.Kharzeev and H.Satz, Phys. Lett. {\bf B356} (1995) 365 , D.Kharzeev and
 H.Satz, Phys. Lett. {\bf B366}(1996) 316
\bibitem{r10} J.Pi\v{s}\'{u}t,N.Pi\v{s}\'{u}tov\'a and P.Z\'avada,
  Zeit.f. Phys. {\bf C67} (1995) 467
\bibitem{r11} R.Lietava,J.Pi\v{s}\'{u}t,N.Pi\v{s}\'{u}tov\'a and P.Z\'avada,
  Strangeness enhancement in proton-nucleus collisions in a parton model,
  Comenius University Bratislava,in preparation; J.Pi\v{s}\'{u}t,talk
  at the WA-97 Meeting,January 1996,CERN
\bibitem{r12} C.-Y.Wong, Phys.Rev.Lett. {\bf 76} (1996) 196
\bibitem{r13} X.-M.Xu,D.Kharzeev,H.Satz and X.-N.Wang, $J/\Psi $
    suppression in an equilibrating Parton Plasma, preprint CERN- TH/95
    -304; hep- ph/ 9511331, Nov.1995
\bibitem{r14} T.Sj\"{o}strand and M. van Zijl,Phys.Rev. {\bf D36} (1987)
 2019; L.Durand and H. Pi, Phys. Rev. {\bf D40} (1989) 1436; N. Abou el Naga,
 K.Geiger and B.M\"{u}ller, J.Phys.G {\bf G18} (1992) 797; M.Borzumati and
 G.Kramer, Zeit. f. Phys. {\bf C67} (1995) 137; G.A. Schuler and
 T.Sj\"{o}strand, Phys.Rev. {\bf D49} (1994) 2257, M.M.Block et al., Phys.
 Rev. {\bf D45} (1992) 839
\bibitem{r15} A.Bialas et al.,Nucl.Phys. {\bf B111} (1976) 461; A.Bialas
 and W.Czyz, Nucl.Phys. {\bf B194} (1982) 21; A.Bialas et al., Phys. Rev.
 {\bf D25} (1982) 2328
\bibitem{r16} H.Kichimi et al., Phys.Rev. {\bf D20} (1979) 37
\bibitem{r17} J.W.Chapman et al., Phys.Lett. {\bf 47B} (1973) 465; M.Alston-
 Garjost et al., Phys.Rev. Lett. {\bf 35} (1975) 142; K.Jaeger et al.,
 Phys.Rev. {\bf D11} (1975) 1756 and 2405; A.Sheng et al., Phys.Rev. {\bf
 D11} (1975) 1733; P.Skubic et al., Phys. Rev. {\bf D18} (1978) 3115
\bibitem{r18} M.Asai et al., Zeit.f.Phys. {\bf C27} (1985) 11
\bibitem{r19} T.Kachelhoffer and W.Geist,Estimates of relative yields
 of strange baryons and antibaryons from pp and pA interactions,
 Strasbourg preprint CRN 96-03
\bibitem{r20} R.C.Hwa,Phys.Rev.Lett. {\bf 52} (1984) 493; J.H\"{u}fner
 and A.Klar, Phys.Lett. {\bf B145} (1984) 167;K.Kinoshita et al., Progr.
 Theor. Phys. {\bf 63} (1980) 928,R.C.Hwa and M.Zahir, Phys.Rev. {\bf D31}
 (1985) 499; S.Dat\'{e} et al., Phys.Rev. {\bf D32} (1985) 619
\bibitem{r21} E.L.Feinberg and I.Ya.Pomeranchuk,Suppl. Nuovo Cim. {\bf 3}
 (1956) 652
\bibitem{r22} F.Karsch et al.,Zeit.f.Phys. {\bf C60} (1993) 519;
     S.Gottlieb et al.,Phys.Rev. {\bf D35} (1987) 3972;R.V.Gavai et al.,
     Phys.Lett. {\bf B241} (1990) 437
\bibitem{r23} J.Dodd,NA-44 Collaboration,Talk at the 25th International
  Symposium on Multiparticle Dynamics,Star\'a Lesn\'a,12th-16th Sept.1995,
  to be published in the Proceedings; S.Slegt et al. WA-93 Collab.,
  Nucl.Phys. {\bf A590} (1995) 469c
\bibitem{r24} T.Alber et al.,NA-35 Collab., Zeit.f. Phys. {\bf C66} (1995)
   77
\bibitem{r25} J.Pi\v{s}\'ut,N.Pi\v{s}\'utov\'a and P.Z\'avada,Comenius
    University ,Bratislava, in preparation
 \bibitem{r26} H.Leutwyler and A.V.Smilga, Nucl.Phys. {\bf B342} (1990) 437
\bibitem{r27} J.Ellis and K.Geiger, Phys.Rev. {\bf D52} (1995) 1500;
   CERN-TH./95- 283 (1995) and hep-ph/ 9511321; CERN- TH./96- 105 and
   hep-ph/ 9605425
\bibitem{r28} B.R.Webber, Nucl.Phys. {\bf B238} (1984) 492; G.Marchesini
    and B.R. Webber, Nucl.Phys. {\bf B349} (1991) 617
\bibitem{r29} S.Gavin and R.Vogt, Charmonium suppression by comover
   scattering in Pb+Pb collisions, preprint LBL- 37980, (1996);
   S. Gavin et al., Zeit. f. Phys. {\bf C61} (1994) 351;
   S.Gavin and R.Vogt, Transverse momentum of $\Psi $ and dimuon in
   Pb+Pb collisions, preprint CU- TP- 791
\bibitem{r30} J.-P.Blaizot and J.- Y. Ollitrault, Phys. Rev. Lett.
   {\bf 77} (1996) 1703
\bibitem{r31} D.Kharzeev, Talk at Quark Matter Conference, Heidelberg,
   May, 1996, hep- ph/ 96o9025
\end{thebibliography}
\end{document}